\documentstyle{article}
\title{Bounds on the heat kernel of the Schr\"odinger operator in a random electromagnetic field }
\author{ Z. Haba\\Institute of Theoretical Physics, University of Wroclaw,
\\50-204 Wroclaw, Plac Maxa Borna 9, Poland\\e-mail:zhab@ift.uni.wroc.pl}
\date{}
\begin{document}
\maketitle
\begin{abstract}
We obtain  lower and upper bounds on the heat kernel and Green
functions of the Schr\"odinger operator in a random Gaussian
magnetic field and a fixed scalar potential. We apply stochastic
Feynman-Kac representation, diamagnetic upper bounds and the
Jensen inequality for the lower bound. We show that if the
covariance of the electromagnetic (vector) potential is increasing
at large distances then the lower bound is decreasing
exponentially fast for large distances and a large
time.\end{abstract}
\section{ Introduction}

Random magnetic fields appear in many models of physical interest.
In the fundamental quantum theory the random magnetic field can be
considered as a thermal (high temperature) part of the quantum
electromagnetic field at finite temperature
\cite{mil}\cite{mandel}. The classical random magnetic field is
discussed in optics \cite{goodman} and in a description of an
interaction of light with atoms \cite{mandel}. A random magnetic
field can arise as a Lagrange multiplier in models of interacting
quantum particles creating four-fermion (or
four-boson)interactions \cite{ioffe}\cite{shankar}. In the
Ginzburg-Landau theory of superconductivity \cite{fetter} when
fluctuations are taken into account then the electromagnetic field
becomes a random Gaussian variable \cite{halperin}.

The effect of a random electric field in models of quantum
mechanics has been well elaborated. Anderson localization
\cite{anderson} has been proved for a large class of models
\cite{pastur}. It seems that much less is known rigorously
concerning the localization properties of the Hamiltonians with a
random magnetic field (see \cite{erdos} for a recent review; a
special case of a magnetic field orthogonal to a plane and varying
inside the plane is discussed in \cite{ueki}\cite{leschke}). Some
aspects of localization (e.g., the integrated density of states
\cite{ueki}-\cite{leschke}) can be studied by means of the heat
kernel of the Schr\"odinger Hamiltonian in a random
electromagnetic field. The simplest estimate on the heat kernel
follows from the diamagnetic inequality \cite{combes}\cite{simon}
which bounds the heat kernel in a magnetic field by that without
the magnetic field. Such estimates are not interesting if we wish
to detect the impact of the magnetic field on physical systems.
Stronger upper bounds on the heat kernel in a (deterministic)
magnetic field have been discussed in \cite{mal}
\cite{erdos2}\cite{loss}\cite{nakamura}\cite{avron}. In these
estimates the contribution of the lowest positive eigenvalues and
eigenfunctions has been estimated.

We discuss a lower bound on the expectation value of the heat
kernel. The heat kernel is gauge dependent. We explain which
properties of the heat kernel do not depend on the choice of the
gauge. Our result admits a fast decay of the heat kernel for large
distances and a large time if the random vector potential has
growing correlation functions. The well-known example of a
constant magnetic field \cite{avron} shows that the exponential
decay is really possible. The effect of the magnetic field upon
the upper bounds is hard to derive.  The classical
Cwickel-Lieb-Rosenbljum  bound  on the number of eigenvalues
\cite{simon} depends on momentum and the vector potential ${\bf
A}$ in the combination $\vert {\bf p}+{\bf A}\vert$. Hence,
 the dependence on the vector potential ${\bf A}$ drops out.
 A more precise bound has been derived in \cite{helffer}-\cite{shen}
 which shows a dependence on the magnetic field. Results discussed
  in \cite{ueki} - \cite{mal} display the discrete
spectrum  in the asymptotic behaviour of the heat kernel for large
time and large distances. Although our lower bound does not give
an exact behaviour of the heat kernel, we nevertheless have a
feeling that the exponentially decreasing lower bounds reflect an
intrinsic property of the growing vector potential. We discuss in
the last section a simple example of a non-trivial bound from
above which supports our argument that growing vector potentials
improve localization.

 The heat kernel defines Green functions in Euclidean quantum
field theory
 of an electromagnetic field interacting with a complex scalar field.
The relation follows from Schwinger's proper time formalism
\cite{schwinger}: the correlation functions of the scalar field
can be calculated from the heat kernel by an integral over the
(proper) time. We consider the Ginzburg-Landau model with the
Lagrangian
\begin{displaymath}
{\cal L}=\frac{1}{2}(i\hbar \nabla +{\bf A})\overline{\phi}
(-i\hbar \nabla + {\bf A})\phi+\frac{1}{2}m^{2}\vert \phi\vert^{2}
+\beta\vert\phi\vert^{4} +\frac{1}{4}F_{jk}F^{jk}\end{displaymath}
where $F_{jk}=\partial_{j}A_{k}-\partial_{k}A_{j}$.

The electromagnetic Lagrangian leads to ultraviolet problems which
should be irrelevant for a large distance behaviour. We regularize
the electromagnetic potentials and  discuss Euclidean scalar
fields in a regular Gaussian electromagnetic field ( the
regularity can be achieved by an addition of higher order
derivatives of the electromagnetic field to the Lagrangian). We
obtain lower and upper bounds on the correlation functions of the
scalar field. Decay of correlation functions of the scalar field
describes a disappearance of the long range correlations of the
order parameter in the Ginzburg-Landau model of superconductivity
\cite{fetter}\cite{halperin}.

\section{The imaginary time evolution}
We can apply simple inequalities only for the imaginary time
evolution
\begin{equation}
\frac{d}{d\tau}\psi={\cal A}\psi
\end{equation}
where $-\hbar{\cal A}({\bf A},V)$ is the Hamiltonian with a random
vector potential ${\bf A}$ and a scalar potential $V$(the scalar
potential could have been random but its eventual randomness would
not change our results concerning random magnetic fields, so we
treat it as deterministic)
\begin{equation}\hbar {\cal A}=-\frac{1}{2m}(-i\hbar \nabla +{\bf
A})^{2}-V
\end{equation}
The imaginary time can be treated as the inverse temperature
$\beta$ of the quantum Gibbs distribution
\begin{displaymath}
\hbar\beta= \tau
\end{displaymath}
In this section we assume that the  scalar potential $V$ is
continuous and bounded from below and the sample functions of the
vector potential ${\bf A}$ are continuously differentiable (almost
surely). In the next section we relax the assumptions on ${\bf A}$
(admitting continuous transverse vector potentials).

 Under our assumptions
 on the potentials for almost every sample path ${\bf A}$
 the solution of eq.(1) defines a random semigroup $T_{\tau}$
 \cite{random} expressed in the
form \cite{simon}\cite{truman}
\begin{equation}\begin{array}{l}
\psi_{\tau}({\bf x})\equiv (T_{\tau}\psi)({\bf x })\cr
=E\Big[\exp\Big(\frac{i}{\hbar}\int_{0}^{\tau}{\bf A}({\bf
x}+\sigma {\bf b}_{s})\circ \sigma d{\bf
b}_{s}-\frac{1}{\hbar}\int_{0}^{\tau}V({\bf x}+\sigma {\bf
b}_{s})ds\Big)\psi({\bf x}+\sigma{\bf b}_{\tau})\Big]\end{array}
\end{equation}
where \begin{equation}\sigma=\sqrt{\frac{\hbar}{m}}\end{equation}
and ${\bf b}$ is the Brownian motion defined as the Gaussian
process with the covariance
\begin{equation}
E[b_{j}(s)b_{k}(s^{\prime})]=\delta_{jk}min(s,s^{\prime})\end{equation}
 The
stochastic integral in eq.(3) is defined in the Stratonovitch
sense \cite{ikeda}. We have \begin{equation}\sigma \int {\bf
A}\circ d{\bf b}(s)=\sigma\int {\bf A}d{\bf b}(s)+\int
\frac{\sigma^{2}}{2}\nabla {\bf A}ds
\end{equation}
where the integral on the rhs is in the Ito sense. Hence, if  the
vector potential is in the transverse gauge then the Stratonovitch
and Ito integrals coincide.

Eq.(3) defines the kernel $K^{({\bf A},V)}_{\tau}(x,y) $ as an
integrable function on $R^{D}\times R^{D}$. Under the assumptions
that $V$ is a continuous function bounded from below and the
sample paths of ${\bf A}$ are continuously differentiable there
exists the unique kernel continuous on $R^{D}\times R^{D}$ and
determined  by the formula \cite{simon} \cite{leschkejfa} (for a
minimal set of assumptions which allow to derive the stochastic
representation of the kernel see \cite{hundert})\begin{equation}
\begin{array}{l}
\displaystyle{K_{\tau}^{({\bf A},V)}({\bf x}^{\prime},{\bf
x})\equiv\exp(\tau {\cal A})({\bf x}^{\prime},{\bf
x})=(2\pi\tau\sigma^{2})^{-\frac{D}{2}}\exp(-\frac{1}{2\tau\sigma^{2}}({\bf
x}-{\bf x}^{\prime})^{2})}\cr \cr
\displaystyle{E\Big[\exp\Big(\frac{i}{\hbar}\int_{0}^{\tau}{\bf
A}({\bf q}(s))\circ d{\bf q}_{s}-\frac{1}{\hbar}\int_{0}^{\tau}V(
{\bf q}_{s})ds\Big)\Big]}
\end{array}
\end{equation}
here
\begin{equation}
{\bf q}_{s}={\bf x}+({\bf x}^{\prime}-{\bf
x})\frac{s}{\tau}+\sigma\sqrt{\tau}{\bf a}(\frac{s}{\tau})
\end{equation}
where the Brownian bridge ${\bf a}$ is the Gaussian process on
$[0,1]$ starting in ${\bf 0}$ at $s=0$ and ending in ${\bf 0}$ at
$s=1$ with the covariance \begin{displaymath}
E[a_{j}(s)a_{k}(s^{\prime})]=\delta_{jk}s^{\prime}(1-s)
\end{displaymath} if $s^{\prime}\leq s$. It can be expressed by the Brownian motion
\begin{displaymath}
{\bf a}(s)=(1-s){\bf b}(\frac{s}{1-s})
\end{displaymath}
From the representation (7) there follows the diamagnetic upper
bound ( consequences of the stochastic representation are
discussed in \cite{combes}\cite{simon}; for an analytic proof see
\cite{schrader})

{\bf Proposition 1}( Kato inequality)

Assume that ${\bf A}$ is a continuously differentiable function
and $V$ is a continuous function bounded from below then the
version of the kernel which is continuous on $R^{D}\times R^{D}$
satisfies the inequality
\begin{equation} \vert K_{\tau}^{({\bf A},V)}({\bf x},{\bf y})\vert
\leq   K_{\tau}^{({\bf 0},V)}({\bf x},{\bf y})
\end{equation}
for any $\tau,{\bf x},{\bf y}$.

 Let us note that if
\begin{equation}{\bf A}^{\prime}={\bf A}+\nabla \chi\end{equation}
then
\begin{equation}
\exp(\tau {\cal A}^{\prime})({\bf x}^{\prime},{\bf x})
 =\exp(\frac{i}{\hbar}\chi({\bf x}^{\prime})-\frac{i}{\hbar}\chi({\bf x}))\exp(\tau {\cal A})({\bf x}^{\prime},{\bf x})
 \end{equation}
 The kernel is not invariant under the gauge transformations.
 However,
 \begin{equation}
 (\psi_{1}^{\prime},\exp(\tau {\cal
 A}^{\prime})\psi^{\prime}_{2})= (\psi_{1},\exp(\tau {\cal
 A})\psi_{2})
\end{equation}
if
\begin{equation}
\psi_{j}^{\prime}=\exp(\frac{i}{\hbar}\chi)\psi_{j}
\end{equation}We consider a random Gaussian electromagnetic field
${\bf A}^{\prime}$  with the
mean equal  zero.  We define the covariance $G^{\prime}$ of ${\bf
A}^{\prime}$ in an arbitrary gauge as the expectation value
\begin{equation}
G_{jk}^{\prime}({\bf x},{\bf x}^{\prime})=\langle
A_{j}^{\prime}({\bf x})A_{k}^{\prime}({\bf x}^{\prime})\rangle
\end{equation}It will be convenient to fix the gauge of the
random vector potential and subsequently discuss a dependence of
our results on the gauge. From eq.(6) it can be seen that the
transverse (Landau) gauge $div {\bf A}=0$ is distinguished from
the point of view of the path  integral (if $div {\bf A}=0$ in a
distributional sense then a continuity of ${\bf A}$ is sufficient
for the stochastic representation (7) of the kernel). If we wish
to calculate the heat kernel in another gauge ${\bf A}^{\prime}$
then we need the gauge transformation with
\begin{equation} \chi= \triangle^{-1}div {\bf A}^{\prime}
\end{equation}
transforming ${\bf A}^{\prime}$ to the transverse gauge. The
covariance $G$ of ${\bf A}$ is related to that of ${\bf
A}^{\prime} $ by the formula
\begin{equation}
G_{jk}({\bf x},{\bf x}^{\prime})= ( \delta_{jr}-
\partial_{j}\partial_{r}\triangle^{-1})( \delta_{km}-
\partial_{k}^{\prime}\partial_{m}^{\prime}\triangle^{\prime -1})G_{rm}^{\prime}({\bf x},{\bf x}^{\prime})
\end{equation}
 We calculate the Gaussian expectation
value of the kernel (7) using the definition of the Gaussian
variable
\begin{equation}
\langle \exp(i({\bf A},{\bf
f}))\rangle=\exp\Big(-\frac{1}{2}\langle ({\bf A},{\bf
f})^{2}\rangle\Big)
\end{equation}
From this formula we can see that a change of the gauge  involves
the factor
\begin{equation}
\exp\left(-\frac{1}{2\hbar^{2}}\left< \left(\chi\left({\bf
x}\right)-\chi\left({\bf x}^{\prime}\right)+\int{\bf
A}^{\prime}\circ d{\bf
q}\right)^{2}\right>\right)=\exp\left(-\frac{1}{2\hbar^{2}}\left<\left(
\int{\bf A} d{\bf q}\right)^{2}\right >\right)\end{equation} in
the kernel (7).

\section{Bounds on the heat kernel and Green functions}
We can calculate the expectation value of the kernel over the
magnetic field
\begin{equation}
\begin{array}{l}
\displaystyle{\langle K_{\tau}^{({\bf A},V)}({\bf x}^{\prime},{\bf
x})\rangle\equiv \langle \exp(\tau{\cal A}({\bf A},V))({\bf
x}^{\prime},{\bf x})\rangle
=(2\pi\tau\sigma^{2})^{-\frac{D}{2}}\exp(-\frac{1}{2\tau\sigma^{2}}({\bf
x}-{\bf x}^{\prime})^{2})}\cr\cr\displaystyle{E\Big[\exp\Big(
-\frac{1}{2\hbar^{2}}\langle (\int_{0}^{\tau}{\bf A}({\bf
q}(s))\circ d{\bf
q}_{s})^{2}\rangle-\frac{1}{\hbar}\int_{0}^{\tau}V({\bf
q}_{s})ds\Big)\Big]}
\end{array}
\end{equation}
 In the formula (19) the Stratonovitch integral
can be expressed by the Ito integral (the integrals coincide for
transverse vector fields). It follows from eq.(19) that
\begin{displaymath}
\langle K_{\tau}^{({\bf A},V)}({\bf x}^{\prime},{\bf
x})\rangle\leq K_{\tau}^{({\bf 0},V)}({\bf x}^{\prime},{\bf x})
\end{displaymath} which is already a consequence of the Kato inequality
(9).

We  rewrite the  Ito integral in eq.(19) in the form ( we can
prove this equality differentiating both sides of eq.(20) below by
means of the Ito formula \cite{ikeda} and subsequently integrating
it again, see also \cite{mizel})
\begin{equation}
(\int_{0}^{\tau}{\bf A}d{\bf q})^{2}=2\int_{0}^{\tau}{\bf A}d{\bf
q}_{s}\int_{0}^{s}{\bf A}d{\bf q}_{s^{\prime}}+
\int_{0}^{\tau}{\bf A}^{2}\frac{ds}{\tau}
\end{equation}
 Let us
note that
\begin{equation}\begin{array}{l} \displaystyle{\int_{0}^{\tau}{\bf A}({\bf
q}(s))\circ d{\bf q}_{s}= \int_{0}^{\tau}{\bf A}({\bf q}(s)) d{\bf
q}_{s}+\frac{\sigma}{2}\int_{0}^{\tau}div{\bf A}({\bf
q}(s))ds}\cr\cr
\displaystyle{=\frac{\sigma}{2}\int_{0}^{\tau}div{\bf A}({\bf
q}(s))ds - \sigma\sqrt{\tau}\int_{0}^{\tau}\frac{ds}{\tau}{\bf
A}({\bf q}(s)){\bf
b}(\frac{s}{\tau-s})}\cr\cr\displaystyle{+\sigma\sqrt{\tau}\int_{0}^{\tau}{\bf
A}({\bf q}(s))
 (1-\frac{s}{\tau}) d{\bf b}(\frac{s}{\tau-s})
-\int_{0}^{\tau}({\bf A}({\bf q}(s))({\bf x}-{\bf
x}^{\prime})\frac{ds}{\tau}}
\end{array}
\end{equation}
consists of four terms which could behave in a different way for
large distances.

  We shall discuss  either a random electromagnetic potential ${\bf A}$
  which
  is bounded (in a certain gauge) or a
random electromagnetic potential   which is scale invariant and
growing with the distance
\begin{equation} {\bf A}(\lambda{\bf x})\simeq
\lambda^{\gamma}{\bf A}({\bf x}) \end{equation} where the
approximate equality means that the random fields on both sides
have the same correlation functions. The scale invariance is
assumed only for a convenience. In general, we can treat  the
scale invariance (22) as an asymptotic behaviour for large
distances.

When $\gamma>0$ then a scale invariant random field in the
transverse gauge (16) must have the covariance (in the Feynman
gauge we could interpret this vector field as $D$-independent
Levy's $D$-dimensional Brownian sheets \cite{levy})
\begin{equation}
G_{jk}({\bf x},{\bf x}^{\prime})= ( \delta_{jr}-
\partial_{j}\partial_{r}\triangle^{-1})( \delta_{kr}-
\partial_{k}^{\prime}\partial_{r}^{\prime}\triangle^{\prime -1})(\vert {\bf x}\vert^{2\gamma}+
\vert {\bf x}^{\prime}\vert^{2\gamma}-\vert {\bf x}-{\bf
x}^{\prime}\vert^{2\gamma})
\end{equation}
with $\gamma<1$. The derivatives on the rhs of eq.(23) should be
understood in the distributional sense. The sample paths of ${\bf
A}$ are H\"older continuous with an index $\alpha$ arbitrarily
close to $\gamma$ \cite{sample}.  A random vector field ${\bf A}$
which is  only H\"older continuous does not satisfy the conditions
for the representation of the kernel formulated in sec.2.
However,the kernel (7) for  ${\bf A}$ in the transverse gauge can
be defined under the weaker condition of H\"older continuity
\cite{hundert}. The Ito stochastic integral in eq.(7) is defined
for any continuous ${\bf A}$. The expectation value (7) determines
a jointly continuous function of $\tau,{\bf x},{\bf y}$ for
continuous potentials ${\bf A}$ and $V$ ( $V$ bounded from below).
Then, the kernel (7) defines a random semigroup $T_{\tau}$
\cite{random} as a result of the the Markov property of the
Brownian motion and the multiplicative property of the
exponentials \cite{simon}. The Hamiltonian  $-\hbar {\cal A}$
(together with its domain) is defined as the generator of
$T_{\tau}$. Alternatively, we could introduce a covariance
$G_{\kappa}$ of the vector potential which is infinitely
differentiable and depending on a regularizing parameter
${\kappa}$ . We consider expectation values of the kernel with a
regularized electromagnetic field. Subsequently, we take the limit
$G_{\kappa}\rightarrow G$ of expectation values (where $G$ is
defined in eq.(23)). The inequalities discussed in this paper are
preserved under such limits.

 We can obtain a
lower bound on the heat kernel from the Jensen inequality
\cite{jen1}-\cite{jen2} as applied in eq.(19) to an average over
the Brownian motion
\begin{equation}
\begin{array}{l}
\displaystyle{\langle\exp(\tau {\cal A})({\bf x}^{\prime},{\bf
x})\rangle\geq(2\pi\tau\sigma^{2})^{-\frac{D}{2}}\exp(-\frac{1}{2\tau\sigma^{2}}({\bf
x}-{\bf x}^{\prime})^{2})}\cr\cr
\displaystyle{\exp\Big(-\frac{1}{2\hbar^{2}}E\Big[\langle
(\int_{0}^{\tau}{\bf A}({\bf q}(s))\circ d{\bf
q}_{s})^{2}\rangle\Big]-\frac{1}{\hbar}\int_{0}^{\tau}V({\bf
q}(s)) ds\Big)}
\end{array}
\end{equation}
On the basis of eq.(24) and the diamagnetic inequality (9)
(Proposition 1) we obtain the following bounds  (the terms on the
lhs of eqs.(25) and (28) below can be related by some inequalities
but we keep this form of the inequalities in order to make  the
origin of these terms visible in such a form as they come from
eq.(24))

{\bf Theorem 2}

Assume  that $c\leq V\leq a$ and the covariance of the
electromagnetic field $\langle A_{j}({\bf x})A_{k}({\bf
x}^{\prime})\rangle$ in the transverse gauge is continuous and
bounded. Then, there exists a constant $C>0$ and positive
constants $a_{j}$ such that
\begin{equation}
\begin{array}{l}
\displaystyle{C(2\pi\tau\sigma^{2})^{-\frac{D}{2}}\exp(-\frac{1}{2\tau\sigma^{2}}({\bf
x}-{\bf x}^{\prime})^{2})}\cr\cr\displaystyle{\exp\Big(
-a_{1}\hbar^{-2}({\bf x}-{\bf
x}^{\prime})^{2}-a_{2}\hbar^{-\frac{3}{2}}\vert{\bf x}-{\bf
x}^{\prime}\vert\sqrt{\tau}-a\hbar^{-1}\tau
\Big)}\cr\cr\displaystyle{\leq\langle K_{\tau}({\bf
x}^{\prime},{\bf x})\rangle \leq \exp
(-\frac{c}{\hbar}\tau)(2\pi\tau\sigma^{2})^{-\frac{D}{2}}\exp(-\frac{1}{2\tau\sigma^{2}}({\bf
x}-{\bf x}^{\prime})^{2})}

\end{array}
\end{equation}

\newpage
{\bf Theorem 3}

i)Assume that  ${\bf A}$ is transverse and continuous and $V$ is
continuous and bounded from below then the kernel satisfies the
upper bound
\begin{equation}
\begin{array}{l}
\displaystyle{\langle K_{\tau}^{({\bf A},V)}({\bf x}^{\prime},{\bf
x})\rangle}\cr\cr
\displaystyle{\leq(2\pi\tau\sigma^{2})^{-\frac{D}{2}}\exp(-\frac{1}{2\tau\sigma^{2}}({\bf
x}-{\bf x}^{\prime})^{2}) \int_{0}^{\tau}\frac{ds}{\tau}
E\Big[\exp\Big( -\frac{\tau}{\hbar}V({\bf q}_{s})\Big)\Big]}\cr\cr
\displaystyle{=
(2\pi\sigma^{2}\tau)^{-\frac{D}{2}}\exp(-\frac{1}{2\tau}({\bf
x}-{\bf x}^{\prime})^{2})}\cr\cr \displaystyle{\int_{0}^{1}ds\int
d{\bf y}(2\pi)^{-\frac{D}{2}} \exp(-\frac{ {\bf
y}^{2}}{2})\exp\Big(-\frac{\tau}{\hbar}V({\bf x}+({\bf
x}^{\prime}-{\bf x})s+\sqrt{\tau}\sigma s(1-s){\bf
y})\Big)}\end{array}
\end{equation}

ii) Assume that the vector potential in the transverse gauge is
scale invariant and growing with the scale index $\gamma$
(eq.(23)), the scalar potential is a continuous function bounded
from below and for certain $a>0$ and $B>0$
\begin{equation}
V({\bf x})\leq B\vert {\bf x}\vert^{2\beta}+a\end{equation}
 then there exists a constant $C>0$ and some positive constants $a_{j}$
 such that

\begin{equation}
\begin{array}{l}

\langle K_{\tau}({\bf x}^{\prime},{\bf x})\rangle\geq
C(2\pi\sigma^{2}\tau)^{-\frac{D}{2}}\exp(-\frac{1}{2\sigma^{2}\tau}({\bf
x}-{\bf x}^{\prime})^{2})\cr\exp\Big( -a_{1}\hbar^{-2}(\vert{\bf
x}\vert^{2\gamma}+\vert{\bf x}^{\prime}\vert^{2\gamma})({\bf
x}-{\bf x}^{\prime})^{2}-a_{2}\hbar^{-\frac{3}{2}}(\vert{\bf
x}\vert^{2\gamma}+\vert{\bf x}^{\prime}\vert^{2\gamma})\vert{\bf
x}-{\bf x}^{\prime}\vert\sqrt{\tau}\cr-a_{3}\hbar^{-1}(\vert{\bf
x}\vert^{2\gamma}+\vert{\bf x}^{\prime}\vert^{2\gamma})\tau
-a_{4}\hbar^{-\frac{3}{2}+\gamma}\tau^{\frac{1}{2}+\gamma}\vert{\bf
x}-{\bf x}^{\prime}\vert-a_{5}{\hbar}^{-1+\gamma}\tau^{1+\gamma}
\cr-a_{6}\hbar^{-2+\gamma}\vert {\bf x}-{\bf
x}^{\prime}\vert^{2}\tau^{\gamma} -a_{7}\hbar^{-1} (\vert{\bf
x}\vert^{2\beta}+\vert{\bf x}^{\prime}\vert^{2\beta})
\tau-a_{8}\hbar^{-1+\beta}\tau^{1+\beta}-a\hbar^{-1}\tau\Big)\end{array}
\end{equation}

{\bf Remarks}:

1.The upper bound (26) does not depend on the vector potential as
a consequence of the Kato inequality (9). The final inequality in
eq.(26) follows from the Jensen inequality as applied to the time
integral.

2. By a scale transformation  (for $V=0$) we could obtain (an
expectation value over a scale invariant magnetic field (23)
)\begin{equation} \langle K^{({\bf A},0)}_{\tau}({\bf
x}^{\prime},{\bf x})\rangle=
(2\pi\sigma^{2}\tau)^{-\frac{D}{2}}\exp(-\frac{1}{2\sigma^{2}\tau}({\bf
x}-{\bf x}^{\prime})^{2})
\exp(-\tau^{1+\gamma}F(\tau^{-\frac{1}{2}}{\bf
x},\tau^{-\frac{1}{2}}{\bf x}^{\prime}))\end{equation} where the
function $F$ has to be determined by an explicit calculation. The
lower bound (28) is in agreement with the scaling (29); it gives
an upper bound for the function $F$.

The heat kernel will depend on the gauge. However, its diagonal
\begin{equation}\begin{array}{l} \langle\exp(\tau {\cal A})({\bf
x},{\bf x})\rangle \geq C(2\pi\tau)^{-\frac{D}{2}}\cr\exp\Big(
-2a_{3}\hbar^{-1}\vert{\bf
x}\vert^{2\gamma}\tau-a_{5}{\hbar}^{-1+\gamma}\tau^{1+\gamma}
 -2a_{7}\hbar^{-1} \vert{\bf
x}\vert^{2\beta}
\tau-a_{8}\hbar^{-1+\beta}\tau^{1+\beta}-a\hbar^{-1}\tau\Big)
\end{array}\end{equation} is gauge invariant.
The diagonal of the heat kernel is equal to the Laplace transform
of the integrated density of states \cite{leschkejfa}\cite{ueki}.
The  integral over the diagonal
\begin{equation}
\langle Tr(\exp(\tau{\cal A}))\rangle =\langle
\sum_{n}\exp\left(-\tau\epsilon_{n}\left({\bf
A},V\right)\right)\rangle=\int d{\bf x}\langle\exp(\tau {\cal
A})({\bf x},{\bf x})\rangle
\end{equation}
is expressing the sum over eigenvalues $\epsilon_{n}({\bf A},V)$
of the Hamiltonian $-\hbar{\cal A}$. We have

{\bf Corollary 4}

For any $\tau>0$ there exists a constant $C>0$ such that
\begin{equation}
\begin{array}{l}C
\tau^{-\nu}\exp(-a_{5}{\hbar}^{-1+\gamma}\tau^{1+\gamma}
-a_{8}\hbar^{-1+\beta}\tau^{1+\beta}-a\hbar^{-1}\tau)\cr \leq
\langle Tr(\exp(\tau{\cal A}))\rangle\leq
(2\pi\tau\sigma^{2})^{-\frac{D}{2}}\int d{\bf
x}\exp(-\frac{\tau}{\hbar} V({\bf x}))\end{array}
\end{equation}
here
\begin{equation}
\nu=\frac{D}{2}(1+\frac{1}{\rho})
\end{equation}
with $\rho=max(\gamma,\beta)$

{\bf Remarks}:

1.The factor $\tau^{-\nu}$ in the lower bound on the lhs of
eq.(32) is non-trivial only for a small time; for large time the
exponential terms decay much faster. If $V=0$ then the index $\nu$
(33) follows already from the scaling (29) ( there is no upper
bound in eq.(32) if $V=0$).

2.The lower bound (32) is a result of the integration over ${\bf
x}$ in eq.(30). The upper bound follows from the bound
(26)\begin{equation}
\begin{array}{l}
\int d{\bf x}\langle K_{\tau}^{({\bf A},V)}({\bf x},{\bf
x})\rangle\leq (2\pi\tau\sigma^{2})^{-\frac{D}{2}}\int d{\bf x}
E\Big[\exp\Big( -\frac{1}{\hbar}\int_{0}^{\tau}V({\bf
x}+\sqrt{\tau}\sigma{\bf
a}_{s})ds\Big)\Big]\cr=(2\pi\tau\sigma^{2})^{-\frac{D}{2}}\int
d{\bf x} \exp\Big( -\frac{\tau}{\hbar} V({\bf x})\Big)
\end{array}
\end{equation}
The upper bound (34) follows from that for $\int d{\bf x} \int
d{\bf x}\langle K_{\tau}^{({\bf A},V)}({\bf x},{\bf x})\rangle$
(Kato inequality) and has been derived earlier in \cite{combes} as
a consequence of the Golden-Thompson inequality.

 3.The lower bound of Corollary 4  is suggesting  that a growing
random vector field has a similar effect as a growing scalar
potential leading to localized states. For  the harmonic
oscillator (with the oscillation frequency $\omega$) $Tr\exp(-\tau
H))=(\sinh (\frac{\omega\tau}{2}))^{-1}$. Hence, an increase of
the index $\nu$ in eq.(33) agrees with the exact formula (the
index $\nu$ has been discussed also in \cite{vasil}). However, the
exponential decrease in the lower bound on the lhs of eq.(32) does
not reflect the exact large time behaviour of the trace of the
heat kernel of the Hamiltonian with a scalar potential.

4.We can obtain the general formula for $\langle K_{\tau}\rangle $
from the transverse case transforming a general potential to the
transverse one and subsequently calculating the average over the
gauge function $\chi$ as in eq.(18).  The behaviour for large
distances would not change substantially in Theorem 2 and Theorem
3 if we worked with an arbitrary gauge. We have assumed the
transverse gauge in order to avoid difficulties with
differentiability of the potentials. Let us explain the problem
using as an example the square of the first term on the rhs of
eq.(21)
\begin{equation}\begin{array}{l}
 \displaystyle{\int_{0}^{\tau}ds\int_{0}^{\tau}ds^{\prime}E[\langle div{\bf
 A}^{\prime}({\bf q}_{s})div{\bf
 A}^{\prime}({\bf q}_{s^{\prime}})\rangle]}\cr\cr
 \displaystyle{=\int_{0}^{\tau}ds\int_{0}^{\tau}ds^{\prime}E[\partial_{j}\partial^{\prime}_{k}
 G_{jk}({\bf q}_{s},{\bf q}_{s^{\prime}})]}
\end{array}\end{equation}
If the second order derivatives of $G$ are bounded then  the term
(35) is bounded by $c\tau^{2}$. However, in eq.(23) (without the
projection on the transverse part) the second order derivative
behaves as $\vert{\bf x}\vert^{2\gamma-2}$ which is singular for
$\gamma<1$. Then, the large distance behaviour will be
$\tau^{2}\vert{\bf x}\vert^{2\gamma-2}=\tau^{1+\gamma}\vert
\tau^{-\frac{1}{2}}{\bf x}\vert^{2\gamma-2}$ in agreement with the
scaling formula (29).

We consider now the Green functions ${\cal G}_{m}$ defined as
solutions of the equation
\begin{equation}( -{\cal A}+\frac{1}{2}m^{2}){\cal G}_{m}=\delta
\end{equation}
The Green function can also be defined as the kernel of the
inverse operator in the Hilbert space of square integrable
functions $L^{2}( d{\bf x})$ \cite{maurin}. Then,
\begin{equation}  (-{\cal A}+\frac{1}{2}m^{2})^{-1}=
\int_{0}^{\infty}d\tau \exp(-\frac{1}{2}m^{2}\tau)\exp(\tau{\cal
A})
\end{equation}
 By
means of an integration over $\tau$ of the diamagnetic inequality
(26) we obtain an upper bound for the Green function in a magnetic
field in terms of the Green function without the magnetic field
\begin{displaymath}
\langle {\cal G}_{m}^{({\bf A},V)}({\bf x}^{\prime},{\bf
x})\rangle \leq{\cal G}_{m}^{({\bf 0},V)}({\bf x}^{\prime},{\bf
x})
\end{displaymath}
Under the assumptions of Theorem 3 we obtain the lower bound
\begin{equation}
 \langle{\cal G}_{m}^{({\bf A},V)}({\bf x},{\bf x}^{\prime})\rangle\geq C\vert{\bf x}-{\bf
 x}^{\prime}\vert^{-D+2}\exp\Big(-\frac{a}{\hbar^{2}}\vert{\bf x}-{\bf
x}^{\prime}\vert^{2}-(\frac{b}{\hbar}+m)\vert{\bf x}-{\bf
x}^{\prime}\vert \Big)\end{equation} The lower bound for the
random magnetic field with the covariance(23) and the scalar
potential $V$ (27) follows from eq.(28) by an integration over
$\tau$. A detailed estimate of the behaviour of such integrals as
a function of $\vert {\bf x}-{\bf x}^{\prime}\vert$ is
complicated. Without detailed estimates we can obtain an
exponential decay in $\vert {\bf x}-{\bf x}^{\prime}\vert$ of the
lower bound for ${\cal G}_{m}({\bf x},{\bf x}^{\prime})$ as
follows from the first $\tau$-independent term in the exponential
on the rhs of eq.(28). It is not clear whether this exponential
decay comes solely from the unprecise lower bound or if it is an
intrinsic property of growing vector potentials.

The Green function is gauge dependent as follows from eq.(11). We
define the diagonal \begin{displaymath} {\cal G}_{m}^{({\bf
A},V)}({\bf x},{\bf x})-{\cal G}_{m}^{({\bf 0},0)}({\bf x},{\bf
x})=\int_{0}^{\infty}d\tau\exp(-\frac{1}{2}m^{2}\tau
)(K_{\tau}^{({\bf A},V)}({\bf x},{\bf x})-K_{\tau}^{({\bf
0},0)}({\bf x},{\bf x}))
\end{displaymath}which is gauge independent (see \cite{pan} for some estimates
close to the diagonal ). From eq.(7)
\begin{equation}\begin{array}{l}\displaystyle{\vert {\cal G}_{m}^{({\bf
A},V)}({\bf x},{\bf x})-{\cal G}_{m}^{({\bf 0},0)}({\bf x},{\bf
x})\vert\leq\int_{0}^{\infty}d\tau\exp(-\frac{1}{2}m^{2}\tau)
(2\pi\tau)^{-\frac{D}{2}}}\cr\cr\displaystyle{
E\Big[\vert\int_{0}^{\tau} {\bf A}({\bf q})\circ d{\bf q}\vert
+\int_{0}^{\tau}ds V({\bf q}_{s})\Big]}
\end{array}\end{equation}
Hence, if $V$ is continuous and  ${\bf A}$ is H\"older continuous
with any index $\gamma>0$ (as in eq.(23)) then the diagonal (39)
is finite if $D\leq 3$. This is so because the $\tau$-integrand on
the rhs of eq.(39) for a small $\tau$ is bounded by
$\tau^{-\frac{D}{2}+\frac{1+\gamma}{2}}$ and for a large $\tau$ it
decays exponentially . In fact, we have $\int A({\bf q})d{\bf
q}=\int ({\bf A}({\bf q})-{\bf A}({\bf x}))d{\bf q}$ (because
${\bf q}$ is a closed path) and for an estimate of the rhs we can
apply eqs.(43)-(46) below ( see similar estimates in \cite{haba}).

Applying the inequality $\vert \langle
F\rangle\vert\leq\langle\vert F\vert\rangle$, the Schwartz
inequality and an estimate of $\langle E[ (\int_{0}^{\tau}{\bf
A}d{\bf q})^{2}]\rangle$ in eq.(41) below we obtain  for $D\leq 3$
and $V\geq 0$
\begin{equation} \vert {\cal G}_{m}^{({\bf 0},0)}({\bf x},{\bf x})-
\langle{\cal G}_{m}^{({\bf A},V)}({\bf x},{\bf x})\rangle\vert
\leq C_{1}(m)\vert {\bf x}\vert^{2\alpha}+C_{2}(m)
\end{equation}
where $2\alpha=max(\gamma,2\beta)$ and $C_{j}(m)>0$.

\section{ The estimates}

 We discuss now  estimates leading to the results of Theorem 2 and Theorem 3.
 The upper bound in eqs.(25)-(26) is an elementary consequence of the formula (19)
 ( or the Kato inequality of the Proposition 1).
  For the lower bound we estimate the expectation value  on the rhs of
 eq.(24). An explicit calculation of the average over the
 electromagnetic field gives
\begin{equation}\begin{array}{l} E[\langle (\int_{0}^{\tau}{\bf A}({\bf
q}(s)) d{\bf q}_{s})^{2}\rangle]=
\displaystyle{\int_{0}^{\tau}\frac{ds}{\tau}\int_{0}^{\tau}\frac{ds^{\prime}}{\tau}}\cr\cr
\displaystyle{({\bf x}-{\bf x}^{\prime})E[G( {\bf q}(s),{\bf
q}(s^{\prime}))]({\bf x}-{\bf x}^{\prime})}\cr\cr +
\displaystyle{\tau\sigma^{2}\int_{0}^{\tau}\int_{0}^{\tau}\frac{ds}{\tau}\frac{ds^{\prime}}{\tau}E[{\bf
b}(\frac{s}{\tau-s})G({\bf q}(s)),{\bf q}(s^{\prime})){\bf
b}(\frac{s^{\prime}}{\tau-s^{\prime}})]} \cr\cr+
\displaystyle{\tau\sigma^{2}\int_{0}^{\tau}
  d(\frac{s}{\tau-s})(1-\frac{s}{\tau})^{2}\sum_{j}E[G_{jj}({\bf q}(s),{\bf
  q}(s))]}
 \cr\cr
 -\displaystyle{2\sigma\sqrt{\tau}\int_{0}^{\tau}\frac{ds}{\tau}\int_{0}^{s}\frac{ds^{\prime}}{\tau}E[{\bf b}(\frac{s}{\tau-s})G( {\bf q}(s),{\bf q}(s^{\prime}))]({\bf
x}-{\bf
x}^{\prime})}\cr\cr-\displaystyle{2\sigma\sqrt{\tau}\int_{0}^{\tau}\frac{ds}{\tau}\int_{0}^{s}\frac{ds^{\prime}}{\tau}E[{\bf
b}(\frac{s^{\prime}}{\tau-s^{\prime}})G( {\bf q}(s),{\bf
q}(s^{\prime}))]({\bf x}-{\bf x}^{\prime})} \cr\cr\displaystyle{
+\sqrt{\tau}\sigma\int_{0}^{\tau}\frac{ds}{\tau} E[({\bf
x}^{\prime}-{\bf x}-\sqrt{\tau}\sigma{\bf
b}(\frac{s}{\tau-s}))\int_{0}^{s}d{\bf
b}(\frac{s^{\prime}}{\tau-s^{\prime}})G({\bf q}(s)),{\bf
q}(s^{\prime}))](1-\frac{s^{\prime}}{\tau})}
\end{array}
\end{equation}
Note that from the possible nine terms in eq.(41) there remained
only six because
\begin{equation}
E[\int_{0}^{\tau}{\bf u}_{s}d{\bf b}(s)]=0
\end{equation}
if ${\bf u}_{s}$ depends  on ${\bf b}(s^{\prime})$ with
$s^{\prime}\leq s$ (non-anticipating integrals \cite{ikeda})

In order to estimate the lower bound in eq.(24) we need some
estimates on the Ito integrals. We have \cite{ikeda}
\begin{equation}
E[(\int {\bf f}d{\bf b}_{s})^{2}]=\int E[{\bf f}^{2}]ds
\end{equation}
Let \begin{equation} F(s)=\int_{0}^{s}{\bf f}(s^{\prime},{\bf
b}(s^{\prime}))d{\bf b}(s^{\prime})\end{equation} then from the
Schwartz inequality
\begin{equation}
\vert E[\int_{0}^{\tau}dsh(s,{\bf b}(s))F(s)]\vert^{2}\leq
\int_{0}^{\tau}ds E[h(s)^{2}] \int_{0}^{\tau}dsE[F(s)^{2}]
\end{equation}and for random ${\bf f}$ and $h$\begin{equation}
\vert E[\langle\int_{0}^{\tau}dsh(s,{\bf
b}(s))F(s)]\rangle\vert^{2}\leq \int_{0}^{\tau}ds\langle
E[h(s)^{2}]\rangle \langle \int_{0}^{\tau}dsE[F(s)^{2}]\rangle
\end{equation}
where from eq.(43)
\begin{equation}
E[F(s)^{2}]=\int_{0}^{s}{\bf f}(s^{\prime},{\bf
b}(s^{\prime}))^{2}ds^{\prime} \end{equation}

For the lower bound in eq.(24) we bound each of the six terms in
eq.(41) from above. Then, we obtain the lower bound for the heat
kernel inserting the upper bound with the minus sign for each term
in the exponential in eq.(24).

{\bf Proof of Theorem 2}:

 The upper bound in Theorem 2 is a direct
consequence of the Kato inequality (Proposition 1). For the lower
bound (25) it is sufficient to insert the upper bound for each
term in eq.(41). The terms without the stochastic integrals in
eq.(41) can be estimated by means of the Schwartz inequality using
the boundedness of $V$ and $G$ whereas for the term with the
stochastic integral we apply the inequalities (45) and (43).

{\bf Proof of Theorem 3}:

We could obtain the lower bound (28) estimating eq.(41). In
particular, we could apply the inequality (45) directly to the
last term in eq.(41) (then $h$ and ${\bf f}$ in eqs.(43)-(45) do
not depend on the electromagnetic field). However, it is
instructive to return to eq.(21) in order   to estimate the
product of the terms in the square of $\int {\bf A}d{\bf q}$
directly. Then, $h$ and $f$ depend linearly on the electromagnetic
field.

We do not estimate all the six terms which come from the square of
$\int{\bf A}d{\bf q}$ in eq.(20)( $div{\bf A}=0$) but concentrate
on three typical terms. The remaining three terms can be estimated
in a similar way. First, let us consider the last term on the rhs
of eq.(20) which comes from the square of the stochastic integral
(20)(it corresponds to the third term on the rhs of eq.(41))
\begin{equation}\begin{array}{l}
\displaystyle{\int_{0}^{\tau} ds\langle E[\langle  {\bf A}({\bf
q}(s)){\bf A}({\bf q}(s))\rangle]}\cr\cr
\displaystyle{=\int_{0}^{\tau} ds\sum_{j}E[G_{jj}({\bf q}(s)),{\bf
q}(s))]}\cr\cr\displaystyle{ \leq 4(D-1)2^{4\gamma}\tau(\vert{\bf
x}\vert^{2\gamma} +\vert{\bf x}^{\prime}\vert^{2\gamma})+
2^{2\gamma}\sigma^{\gamma}\tau^{1+\gamma}\int_{0}^{1}E[\vert{\bf
a}(s)\vert^{2\gamma}]}
\end{array}
\end{equation}
In eq.(48) the H\"older inequality
\begin{equation}
\vert {\bf a}+{\bf b}\vert^{2\gamma}\leq 2^{2\gamma-1}(\vert {\bf
a}\vert^{2\gamma}+\vert {\bf b}\vert^{2\gamma} )\end{equation} has
been applied to the covariance $G$ of eq.(23) with ${\bf q}(s)$
defined in eq.(8). The square of the last term of eq.(21) can be
estimated as follows

\begin{equation}\begin{array}{l}
\displaystyle{\int_{0}^{\tau}
\frac{ds}{\tau}\int_{0}^{s}\frac{ds^{\prime}}{\tau}E[\langle {\bf
A}({\bf q}(s))({\bf x}^{\prime}-{\bf x}){\bf A}({\bf
q}(s^{\prime}))({\bf x}^{\prime}-{\bf x})\rangle]}\cr\cr
\displaystyle{=\int_{0}^{\tau}
\frac{ds}{\tau}\int_{0}^{s}\frac{ds^{\prime}}{\tau}E[({\bf
x}^{\prime}-{\bf x})G({\bf q}(s)),{\bf q}(s^{\prime}))({\bf
x}^{\prime}-{\bf x})]}\cr\cr\displaystyle{\leq a_{1}(\vert{\bf
x}\vert^{2\gamma}+\vert{\bf x}^{\prime}\vert^{2\gamma})({\bf
x}-{\bf x}^{\prime})^{2}+ a_{6}\vert {\bf x}-{\bf
x}^{\prime}\vert^{2}\sigma^{\gamma}\tau^{\gamma}}
\end{array}
\end{equation}
where for an estimate of $G({\bf q}(s)),{\bf q}(s^{\prime}))$ the
same set of inequalities has been applied as in eq.(48).

 Next, let
us consider a term in the square of $\int {\bf A}d{\bf q}$ which
is of the form of the expression appearing on the lhs of eq.(46)
(cf. eqs.(20)-(21) to see where this term comes
from)\begin{equation}
\begin{array}{l}
I=\sigma^{2}\tau\langle E[\int_{0}^{\tau}\frac{ds}{\tau}{\bf
A}({\bf q}(s)){\bf b}(\frac{s}{\tau-s})\int_{0}^{s}{\bf A}({\bf
q}(s^{\prime}))(1-\frac{s^{\prime}}{\tau})d{\bf
b}(\frac{s^{\prime}}{\tau-s^{\prime}})]\rangle
\end{array}\end{equation}
Now \begin{equation}h(s,{\bf b}(s))={\bf A}({\bf q}(s)){\bf
b}(\frac{s}{\tau-s})
\end{equation}
and
\begin{equation}
{\bf f}\left(s^{\prime},{\bf
b}\left(\frac{s^{\prime}}{\tau-s^{\prime}}\right)\right)= {\bf
A}\left({\bf
q}\left(s^{\prime}\right)\right)(1-\frac{s^{\prime}}{\tau})\end{equation}
Therefore
 in eq.(46)
\begin{equation}
\langle E[h^{2}]\rangle =E[{\bf b}(\frac{s}{\tau-s})G({\bf
q}(s)),{\bf q}(s)){\bf b}(\frac{s}{\tau-s})]\end{equation} and
\begin{equation}\begin{array}{l}\displaystyle{\int_{0}^{\tau}ds\int_{0}^{s}ds^{\prime}
\left< E\left[{\bf f}^{2}\left(s^{\prime},{\bf
b}\left(\frac{s^{\prime}}{\tau-s^{\prime}}\right)\right)\right]\right>}
\cr\cr
\displaystyle{=\int_{0}^{\tau}ds\int_{0}^{s}ds^{\prime}(1-\frac{s^{\prime}}{\tau})^{2}
\sum_{j}E[G_{jj}({\bf q}(s^{\prime}),{\bf q}(s^{\prime}))]}
\end{array}\end{equation} Hence, from eqs.(45)-(46) and the inequality
(49) we obtain an estimate on $I$ \begin{equation} \vert
I\vert\leq a_{3}\tau(\vert{\bf x}\vert^{2\gamma}+\vert {\bf
x}^{\prime}\vert^{ 2\gamma}) +a_{5}\tau^{1+\gamma}
\end{equation}
On the basis of eq.(41) and the estimates (48)-(56) it is clear
how the lower bounds in eqs.(25) and (28) come out ( the estimate
in the lower bound (28) on the potential $V$ (27) is a simple
consequence of the inequalities (49) and (24)).

 \section{Discussion and Outlook}
 First of all, let us point out that a deterministic linearly rising vector potential really can lead
  to exponentially decaying
 Green functions.

  The heat kernel of the Hamiltonian with  a constant magnetic field ${\bf B}$
in $D=3$ satisfies the inequality
\cite{avron}\cite{simon}(transverse gauge, the $z$ axis is in the
direction of ${\bf B}$ and ${\bf x}=(x,y,z)$)
\begin{displaymath}
\begin{array}{l}
\vert K_{\tau}(x, y,z;x^{\prime}, y^{\prime},z^{\prime})\vert=
(2\pi\tau)^{-\frac{1}{2}}B(2\pi\sinh B\frac{\tau}{2})^{-1} \cr
\vert\exp\Big(-\frac{1}{2\tau}B^{2}(z-z^{\prime})^{2}-\frac{B^{3}}{4}
\coth (B\frac{\tau}{2})((x-x^{\prime})^{2}+(y-y^{\prime})^{2})
+\frac{iB}{2}(xy^{\prime}-x^{\prime}y)\Big)\vert\cr \leq
(2\pi\tau)^{-\frac{1}{2}}B(4\pi\sinh (B\frac{\tau}{2}))^{-1} \cr
\exp\Big(-\frac{1}{2\tau}B^{2}(z-z^{\prime})^{2}-\frac{B^{3}}{4}
((x-x^{\prime})^{2}+(y-y^{\prime})^{2})\Big)
\end{array}
\end{displaymath}
as\begin{displaymath} \coth (\frac{B\tau}{2})\geq 1
\end{displaymath}

By an integration over $\tau$ (37) we obtain an upper bound on the
Green function
\begin{equation}
\begin{array}{l}
\vert {\cal G}_{m}(x,y,z;x^{\prime},y^{\prime},z^{\prime})\vert\cr
\leq C\exp\Big(
-\frac{B^{3}}{4}((x-x^{\prime})^{2}+(y-y^{\prime})^{2})-B^{\frac{3}{2}}\vert
z-z^{\prime}\vert-m\vert{\bf x}-{\bf
x}^{\prime}\vert\Big)\end{array}
\end{equation}
( in the $\tau$-integral the inequality $\sinh
\frac{B\tau}{2}\leq\frac{1}{2}\exp(\frac{B\tau}{2})$ has been
applied).

The decay of the Green function (57) supports a heuristic argument
that the term ${\bf A}^{2}$ in the Hamiltonian (2) is acting like
a potential (see \cite{kibble} and the precise results of
\cite{helffer}\cite{shen}). Note that the diagonal
\begin{equation} K_{\tau}(x, y,z;x, y,z)=F(\tau)\end{equation}
being independent of any spatial coordinate is not an integrable
function in any of the components of ${\bf x}$.

 From the formula
for the heat kernel in terms of eigenfunctions and eigenvalues one
can study their dependence on the random magnetic field. For this
purpose we would need an upper bound for the heat kernel which is
stronger than the diamagnetic inequality (26) of Theorem 3. We are
unable to derive such estimates in general. In order to study the
localization effects of a random magnetic field we investigate a
particular model. Let us assume that the magnetic field depends
only on coordinates $(x,y)$ of the $XY$ plane. Then, we can choose
${\bf A}=(0,0,A_{3}(x,y))$. In such a case the Hamiltonian
$-\hbar{\cal A}$ reads
\begin{equation}
\hbar{\cal
A}=-\frac{1}{2m}(p+A_{3})^{2}-V_{3}(z)+\frac{\hbar^{2}}{2m}\triangle_{xy}
-V_{2}(x,y)\end{equation}where $\triangle_{xy}$ is the
two-dimensional Laplacian. In eq.(59) we added a potential
$V_{3}(z)$ ensuring a localization in $z$. We investigate the
conditions on $V_{2}$ which imply a finite trace of
$\langle\exp\tau {\cal A}\rangle$. We apply the Golden-Thompson
inequality \cite{golden}\cite{thompson} ( for a precise
formulation and the assumptions see \cite{ruskai})
\begin{equation}
Tr(\exp(\tau {\cal A}))\leq Tr\Big(\exp(\frac{\tau}{2}{\cal
B})\exp(\tau {\cal C})\exp(\frac{\tau}{2}{\cal B})\Big)
\end{equation}
if ${\cal A}={\cal B}+{\cal C}$. The rhs of the inequality (32) is
a consequence of the Golden-Thompson inequality. Now, we choose
 \begin{equation}
\hbar{\cal C}=-\frac{1}{4m}(p+A_{3})^{2}-V_{3}(z)
-V_{2}(x,y)\end{equation}and \begin{equation}\hbar {\cal
B}=-\frac{1}{4m}(p+A_{3})^{2}+\frac{\hbar^{2}}{2m}\triangle_{xy}
\end{equation} We have
\begin{equation} \begin{array}{l}\displaystyle{\exp(\tau {\cal B})({\bf x};{\bf x}^{\prime})
= (2\pi\sigma^{2} \tau)^{-1} \exp(-\frac{1}{2\tau\sigma^{2}}( x-
x^{\prime})^{2}-\frac{1}{2\tau\sigma^{2}}( y- y^{\prime})^{2})}
 \cr\cr\displaystyle{\int dp\exp(\frac{i}{\hbar}p(z^{\prime}-z))E\Big[\exp\Big(
-\frac{1}{4\hbar m}\int_{0}^{\tau}(p+A_{3}({\bf
q}(s)))^{2}ds\Big)\Big]}
\end{array}\end{equation}
here ${\bf q}=(q_{1},q_{2})$ is two dimensional (the components
defined in eq.(8))

\begin{equation}
 \begin{array}{l}\exp(\tau {\cal C})(x, y,z;x^{\prime},
y^{\prime},z^{\prime})= (\pi\sigma^{2} \tau)^{-\frac{1}{2}}
\exp(-\frac{1}{\tau\sigma^{2}}(z-z^{\prime})^{2})
\delta(x-x^{\prime})\delta(y-y^{\prime})
 \cr\exp(-\frac{\tau}{\hbar}V_{2}(x,y))\exp(\frac{i}{\hbar}(z^{\prime}-z)A_{3}(x,y)) E\Big[\exp\Big(
-\frac{1}{\hbar }\int_{0}^{\tau}V_{3}(q_{3}(s))ds\Big)\Big]
\end{array}\end{equation}
where
\begin{displaymath}
q_{3}(s)=z+(z^{\prime}-z)\frac{s}{\tau}+\sqrt{\frac{\tau}{2}}\sigma
a_{3}(\frac{s}{\tau})
\end{displaymath}
From the Golden-Thompson inequality
\begin{equation}
\langle Tr(\exp(\tau {\cal A}))\rangle\leq \int d {\bf x}\int
d{\bf x}^{\prime}\langle\exp(\tau{\cal B})({\bf x},{\bf
x}^{\prime})\exp(\tau{\cal C})({\bf x}^{\prime},{\bf x})\rangle
\end{equation}
The expectation value over the magnetic field on the rhs of
eq.(65) can explicitly be calculated. We perform the calculations
in a special case when the potential $V_{3}$ is quadratic
\begin{equation}
V_{3}(z)=m\omega^{2}z^{2}
\end{equation}
In such a case the expectation value over $q_{3}$ gives the heat
kernel of the harmonic oscillator. After a calculation of
integrals over $z$ and $z^{\prime}$ we obtain
\begin{equation}\begin{array}{l}\displaystyle{\langle Tr(\exp(\tau {\cal A}))\rangle
 \leq(2\pi\tau\sigma^{2})^{-\frac{1}{2}}(\sinh(\omega\tau))^{-1}\int dxdy
 \exp(-\frac{\tau}{\hbar} V_{2}(x,y))}\cr\cr
 \int dp
\langle\exp\left(-\frac{1}{2m\omega}\sinh\left(\omega\tau\right)\left(\cosh\left(\omega\tau\right)+1\right)^{-1}
\left(p+A_{3}\left(x,y\right)\right)^{2}\right)
  \cr\cr\displaystyle{ E\Big[\exp\Big(
-\frac{1}{4\hbar m}\int_{0}^{\tau}(p+A_{3}({\bf
q}(s)))^{2}ds\Big)\Big]\rangle}\cr\cr\displaystyle{\leq
(2\pi\tau\sigma^{2})^{-\frac{1}{2}}(\sinh(\omega\tau))^{-1}\int
dxdy
 \exp(-\frac{\tau}{\hbar} V_{2}(x,y))}
 \cr\cr\displaystyle{ \int dp\int_{0}^{\tau}\frac{ds}{\tau}
\langle
E\Big[\exp\Big(-\frac{1}{2m\hbar\omega}\sinh(\omega\tau)(\cosh(\omega\tau)+1)^{-1}
(p+A_{3}(x,y))^{2}}\cr\cr \displaystyle{ -\frac{\tau}{4\hbar
m}(p+A_{3}({\bf q}(s)))^{2}\Big)\Big]\rangle}
\end{array}\end{equation}

 The
expectation value over the magnetic field on the rhs
 of eq.(67) can be calculated with the result (for the covariance
 (23))
 \begin{equation}\begin{array}{l}
\displaystyle{ \langle Tr(\exp(\tau {\cal A}))\rangle
 \leq C_{1}(2\pi\tau\sigma^{2})^{-1}(\sinh(\omega\tau))^{-1}
 \int dxdy\exp(-\frac{\tau}{\hbar} V_{2}((x,y))}\cr\cr\displaystyle{\int_{0}^{\tau}\frac{ds}{\tau}
 E\Big[\Big(G((x,y),(x,y))+G({\bf q}_{s},{\bf
 q}_{s})\Big)^{-\frac{1}{2}}\Big]}
\cr \cr
 \displaystyle{\leq C_{2}(2\pi\tau\sigma^{2})^{-1}(\sinh(\omega\tau))^{-1}\int dxdy\exp(-\tau
 V_{2}(x,y))(x^{2}+y^{2})^{-\frac{\gamma}{2}}}
\end{array}\end{equation}
Eq.(68) shows that the growing random electromagnetic field
improves localization. As an example we consider
\begin{equation}
V_{2}=\vert x\vert^{\alpha}\vert y\vert^{\alpha}
\end{equation}
(the case $\alpha=2$ has been discussed by Simon  \cite{simon2}).
The classical criterion for a discrete spectrum ( eq.(68) with
$\gamma=0$) is not satisfied ( the region in the phase space with
the classical energy less than $E$ has an infinite volume, see
\cite{simon},\cite{simon2}). However, any $\gamma>0$ (random
vector field with a growing covariance) leads to a finite trace .
Note that the results of \cite{helffer}-\cite{shen} concerning the
discrete spectrum do not apply directly to the vector potential
(23)and the scalar potential (69) because the covariance of the
magnetic field ${\bf B}$ is decaying as $\vert {\bf
x}\vert^{2\gamma -2}$ ($\gamma<1$). Hence,it is bounded in the
mean.

In the model (59) (with $V_{3}=0$) we can obtain some estimates on
the off-diagonal of the heat kernel as well. Let
$\tilde{K}_{\tau}(p;x,y,;x^{\prime},y^{\prime})$ be the Fourier
transform of $K_{\tau}(x,y,z;x^{\prime},y^{\prime},z^{\prime})$ in
$z^{\prime}-z$  then
\begin{equation} \begin{array}{l}\displaystyle{\tilde{K}_{\tau}(p;x,y,;x^{\prime},y^{\prime})
= (2\pi\sigma^{2} \tau)^{-1} \exp(-\frac{1}{2\tau\sigma^{2}}( x-
x^{\prime})^{2}-\frac{1}{2\tau\sigma^{2}}( y- y^{\prime})^{2})}
\cr\cr\displaystyle{ E\Big[\exp\Big(
-\frac{1}{\hbar}\int_{0}^{\tau}(\frac{1}{2 m}(p+A_{3}({\bf
q}(s)))^{2}+V_{2}({\bf q}(s)))ds\Big)\Big]}\cr \cr\displaystyle{
\leq (2\pi\sigma^{2} \tau)^{-1} \exp(-\frac{1}{2\tau\sigma^{2}}(
x- x^{\prime})^{2}-\frac{1}{2\tau\sigma^{2}}( y- y^{\prime})^{2})}
 \cr\cr\displaystyle{\int_{0}^{\tau}\frac{ds}{\tau}E\Big[\exp\Big(
-\frac{\tau}{2\hbar m}(p+A_{3}({\bf q}(s)))^{2}
-\frac{\tau}{\hbar}V_{2}({\bf q}(s)) \Big)\Big]}
\end{array}\end{equation}
The expectation value over $A_{3}$ can be calculated  exactly. Let
us consider a simple case of  a translation invariant Gaussian
field with $G({\bf x},{\bf x}^{\prime})=G({\bf x}-{\bf
x}^{\prime})$ (there is no scale invariance if $G({\bf 0})$ is
finite). After a calculation of the expectation value on the rhs
of eq.(70) we obtain
\begin{equation} \begin{array}{l}\displaystyle{\langle\tilde{K}_{\tau}(p;x,y,;x^{\prime},y^{\prime})
\rangle   \leq (2\pi\sigma^{2} \tau)^{-1}
\exp(-\frac{1}{2\tau\sigma^{2}}( x-
x^{\prime})^{2}-\frac{1}{2\tau\sigma^{2}}( y- y^{\prime})^{2})}
 \cr\cr\displaystyle{\int_{0}^{\tau}\frac{ds}{\tau}E\Big[\exp\Big(
-\frac{1}{2}p^{2}(\frac{m\hbar}{\tau}+G(0))^{-1}
-\frac{\tau}{\hbar}V_{2}({\bf q}(s))
\Big)\Big](1+\frac{\tau}{m\hbar}G(0))^{-\frac{1}{2}}}
\end{array}\end{equation}
When $\tau \rightarrow \infty$ then  the upper bound of eq.(71) is
decreasing as $\exp(-\frac{1}{2G(0)}p^{2})$ for a large $p$. We
could interpret such a decay of the Fourier transform of the heat
kernel as a confirmation of the behaviour
$\exp(-a_{1}\hbar^{-2}(z-z^{\prime})^{2})$ (which has also a
Gaussian Fourier transform) in the lower bound (25) of Theorem 2
(scale invariance of the vector potential has not been assumed
there).

 The decay of Green
functions is important for  correlation functions of the complex
scalar fields interacting with an electromagnetic field in the
Ginzburg-Landau model
\begin{equation}
\langle \phi^{*}({\bf x})\phi({\bf x}^{\prime})\rangle =\langle
{\cal G}_{m}^{({\bf A},0)}({\bf x},{\bf x}^{\prime})\rangle
\end{equation}
where  ${\cal G}_{m}^{({\bf A},0)}$ is defined in eq.(37).

The lower and upper bounds on the higher order correlations of the
scalar fields can be studied by means of our methods as well. In
such a case the integral $\int {\bf A}d{\bf q}$ must be extended
to many paths joining the points ${\bf x}_{j}$ as the arguments of
the scalar fields $\phi$. For example
\begin{equation}\begin{array}{l}
\displaystyle{\langle \phi^{*}({\bf x})\phi^{*}({\bf y})\phi({\bf
y}^{\prime})\phi({\bf x}^{\prime})\rangle =\langle {\cal G}^{({\bf
A},0)}({\bf x},{\bf x}^{\prime}){\cal G}^{({\bf A},0)}({\bf
y},{\bf y}^{\prime})\rangle +({\bf x}\rightarrow {\bf y})}\cr\cr
\displaystyle{=\int d\tau
d\tau^{\prime}(2\pi\tau\sigma^{2})^{-\frac{D}{2}}
(2\pi\tau^{\prime}\sigma^{2})^{-\frac{D}{2}}\exp(-\frac{1}{2\tau^{\prime}\sigma^{2}}({\bf
y}-{\bf y}^{\prime})^{2})-\frac{1}{2\tau\sigma^{2}}({\bf x}-{\bf
x}^{\prime})^{2}) }\cr\cr\displaystyle{ \langle
E\Big[\exp\Big(\frac{i}{\hbar}\int_{0}^{\tau^{\prime}}{\bf A}({\bf
q}_{{\bf y}{\bf y}^{\prime}})\circ d{\bf q}_{{\bf y}{\bf
y}^{\prime}}+\frac{i}{\hbar}\int_{0}^{\tau}{\bf A}({\bf q}_{{\bf
x}{\bf x}^{\prime}})\circ d{\bf q}_{{\bf x}{\bf
x}^{\prime}}\Big)\Big]\rangle+({\bf x}\rightarrow {\bf
y})}\end{array}\end{equation}where $({\bf x}\rightarrow {\bf y})$
means the same expression but with exchanged arguments. We can
calculate the expectation value over the electromagnetic field and
derive upper and lower bounds for the correlation functions (73).
The important question to be answered is whether the decay of
correlations holds true for any two points tending to infinity in
the multi-point correlation functions of the scalar fields. This
problem needs further investigation.


\begin{thebibliography}{99}
\bibitem{mil}P.W. Milonni, The Quantum Vacuum, Academic, New
York,1994
\bibitem{mandel}L. Mandel and E. Wolf, Optical Coherence and
Quantum Optics,Cambridge Univ.Press,1995
\bibitem{goodman}J.W. Goodman, Statistical Optics,Wiley,New
York,1985
\bibitem{ioffe}L.B. Ioffe and P.B. Wiegmann, Phys.Rev.Lett.{\bf
65},653(1990)
\bibitem{shankar} G. Murthy and R.Shankar,Rev.Mod.Phys.{\bf
75},1101(2003)

\bibitem{fetter}A.L.Fetter and J.D. Walecka, Quantum Theory of
Many-Particle Systems, McGraw-Hill,1971
\bibitem{halperin}B.I. Halperin, T.C. Lubensky and S.-K. Ma,
Phys.Rev.Lett.{\bf 32},292(1974)
\bibitem{anderson}P.W. Anderson,
Phys.Rev.{\bf 109},1492(1958)

\bibitem{pastur} L. Pastur
and A. Figotin, Spectra of Random and Almost-Periodic
Operators,Springer, Berlin,1992
\bibitem{erdos}L. Erd\"os, arxiv:math-ph/0510055
\bibitem{ueki}N. Ueki,Ann.Inst.Henri Poincare, {\bf 1},473(2000)
\bibitem{leschke}H. Leschke,S. Warzel and A.
Weichlein,arxiv:math-ph/0507035
\bibitem{combes}J.M Combes, R. Schrader and R.Seiler,
Ann.Phys.{\bf 111},1(1978)
 \bibitem{simon}
 B. Simon, Functional Integration and Quantum Physics,
 Academic Press, 1979

 \bibitem{mal} P. Malliavin,C.R.Acad.Sci.Paris Ser.I.Math.{\bf
302},481(1986)\bibitem{erdos2}L. Erd\"os, Duke Math.Journ.{\bf
76},541(1994)
\bibitem{loss} M. Loss and B. Thaller, Commun.Math.Phys.{\bf
186},95(1997)

\bibitem{nakamura} S. Nakamura,Commun.Math.Phys.{\bf
214},565(2000)
\bibitem{avron}
J.E. Avron, I.Herbst and B.Simon, Duke Math.Journ. {\bf
45},847(1978)
\bibitem{helffer}
B. Helffer and A. Mohamed, Ann.l'inst. Fourier,{\bf 38},95(1988)
\bibitem{shen}Z. Shen,Trans.Amer.Math.Soc.{\bf 348},4465(1996)
 \bibitem{schwinger} J.Schwinger,
Phys.Rev.{\bf 82},664(1951) \bibitem{random} A.V. Skorohod, Random
Linear Operators, Reidel, Dordrecht,1984\bibitem{truman}D.K.
Elworthy, A. Truman and K. Watling,
 J.Math.Phys.{\bf 26},984(1985)\bibitem{leschkejfa}K.Broderix, H. Leschke and P.
M\"uller, Journal Funct.Anal.{\bf
212},287(2004)
\bibitem{hundert}K. Broderix, D. Hundertmark and H.
Leschke,
\newline
Rev. Math.Phys.{\bf 12},181(2000)
\bibitem{sample}R.M. Dudley, Ann.Probab.{\bf 1},66(1973)

X. Fernique,in  Lect.Notes.in Math.{\bf
480},Berlin,Springer,1975\bibitem{schrader} H. Hess,R. Schrader
and D. Uhlenbrock, J.Diff.Geom.{\bf 15},27(1980)



\bibitem{ikeda} N. Ikeda and S. Watanabe, Stochastic
Differential Equations and Diffusion Processes, North Holland,1981

\bibitem{mizel}M.A. Berger and V.J.Mizel,
Trans.Amer.Math.Soc.{\bf 252},249(1979)
\bibitem{levy}P. Levy, Processus Stochastiques et Mouvement
Brownien, deuxieme edition, Revue et Augmentee,Paris,1965


 \bibitem{jen1}
J.L.W. Jensen, Acta Math.{\bf 30},175(1906)
\bibitem{jen2}
A. W. Marshall and I. Olkin, Inequalities:Theory of Majorization
and Its Applications, Academic Press,1979

\bibitem{vasil}D.V.Vassilevich, Phys.Rep.{\bf 388},279(2003)
\bibitem{maurin} K. Maurin, Methods of Hilbert
Spaces,PWN,Warszawa,1972



\bibitem{pan}J.Br\"uning, V.Geyler
and K.Pankrashkin,  J.Math.Phys.{\bf 46},113508(2005)

\bibitem{haba}Z.Haba, Phys.Rev.{\bf D26},3506(1982)
\bibitem{golden}S. Golden, Phys.Rev.{\bf 137},B1127(1965)
\bibitem{thompson}C.J. Thompson, J.Math.Phys.{\bf 6},1812(1965)
\bibitem{ruskai}M.B. Ruskai,Commun.Math.Phys.{\bf 26},280(1972)
\bibitem{kibble}T.W.B.Kibble, Phys.Rev.{\bf 150},1060(1966)

\bibitem{simon2}B. Simon, Ann.Phys.{\bf 146},209(1983)




\end{thebibliography}
\end{document}